# ANI-1: An extensible neural network potential with DFT accuracy at force field computational cost

Justin S. Smith[1], Olexandr Isayev[2,*], Adrian E. Roitberg[1,*]

[1]*Department of Chemistry, University of Florida, Gainesville, FL 32611, USA*

[2]*UNC Eshelman School of Pharmacy, University of North Carolina at Chapel Hill, Chapel Hill, NC 27599, USA*

* Corresponding authors; email: O.I. (olexandr@olexandrisayev.com) or A.E.R. (roitberg@ufl.edu)

## Abstract

Deep learning is revolutionizing many areas of science and technology, especially image, text, and speech recognition. In this paper, we demonstrate how a deep neural network (NN) trained on quantum mechanical (QM) DFT calculations can learn an accurate and transferable potential for organic molecules. We introduce ANAKIN-ME (Accurate NeurAl networK engINe for Molecular Energies) or ANI in short. ANI is a new method designed with the intent of developing transferable neural network potentials that utilize a highly-modified version of the Behler and Parrinello symmetry functions to build single-atom atomic environment vectors (AEV) as a molecular representation. AEVs provide the ability to train neural networks to data that spans both configurational and conformational space, a feat not previously accomplished on this scale. We utilize ANI to build a potential called ANI-1, which was trained on a subset of the GDB databases with up to 8 heavy atoms to predict total energies for organic molecules containing four atom types: H, C, N, and O. To obtain an accelerated but physically relevant sampling of molecular potential surfaces, we also propose a Normal Mode Sampling (NMS) method for generating molecular conformations. Through a series of case studies, we show that ANI-1 is chemically accurate compared to reference DFT calculations on much larger molecular systems (up to 54 atoms) than those included in the training data set.

# 1 Introduction

Understanding the energetics of large molecules plays a central role in the study of chemical and biological systems. However, because of extreme computational cost, theoretical studies of these complex systems are often limited to the use of approximate methods, compromising accuracy, in exchange for a speedup in calculations. One of the grand challenges in modern theoretical chemistry has been to design and implement approximations that expedite *ab-initio* methods without loss of accuracy. Popular strategies include the partition of the system of interest into fragments[1,2], linear scaling[3], semi-empirical[4–6] (SE) methods or the construction of empirical potentials that have been parametrized to reproduce experimental or accurate *ab-initio* data.

In SE methods, some of the computationally expensive integrals are replaced with empirically determined parameters. This results in a very large speed up. However, the accuracy is also substantially degraded compared to high level *ab-initio* methods due to imposed approximations.[7] Also, the computational cost of SE is still very high compared to classical force fields (FF), potentially limiting the system size that can be studied.

Classical force fields or empirical interatomic potentials (EP) simplify the description of interatomic interactions even further by summing components of the bonded, angular, dihedral, and non-bonded contributions fitted to a simple analytical form. EP can be used in large-scale atomistic simulations with significantly reduced computational cost. More accurate EPs have been long sought after to improve statistical sampling and accuracy of molecular dynamics (MD) and Monte-Carlo (MC) simulations. However, EP are generally reliable only near equilibrium. These, typically *nonreactive* empirical potentials, are widely used for drug design, condensed matter and polymer research.[8–11] Thus, such potentials are usually not applicable for investigations of chemical reactions and transition states. One exception to this is the ReaxFF force field[12], which is capable of studying chemical reactions and transition states. However, ReaxFF, like most reactive force fields, must generally be reparametrized from system to system and therefore lacks an "out-of-the-box" level of transferability. Furthermore, each application of FF and EP needs to be carefully pondered, as their accuracy varies among different systems. In fact, performing benchmarks to determine the optimal FF combination for the problem at hand is usually unavoidable. Unfortunately, there are no systematic ways for improving or estimating the transferability of EPs.

Machine learning (ML) is emerging as a powerful approach to construct various forms of transferrable[13–15] and non-transferable[16,17] atomistic potentials utilizing regression algorithms. ML methods have been successfully applied in a variety of applications in chemistry, including the prediction of reaction pathways[18], QM excited state energies[19], formation energies[20], atomic forces and nuclear magnetic resonance chemical shifts[21], and assisting in the search of novel materials[22]. ML potentials have shown promise in predicting molecular energies with QM accuracy with as much as 5 orders of magnitude speed up. The key to the transferable methods is finding a correct molecular representation that allows and improves learning in the chosen ML method. As discussed by Behler[23], there are three criteria that such representations must adhere to in order to ensure energy conservation and be useful for ML models: they must be rotationally and translationally invariant, the exchange of two identical atoms must yield the same result, and given a set of atomic positions and types the representation must describe a molecule's conformation in a unique way. Several such representations have been developed[24–27], but true transferability and extensibility to complex chemical environments, i.e. all degrees of freedom for arbitrary organic molecules, with chemical accuracy has yet to be accomplished.

In 2007, Behler and Parrinello (BP) developed an approximate molecular representation, called symmetry functions (SF), that take advantage of chemical locality in order to make neural network potentials[25] (NNP) transferable. These SFs have been successfully applied to chemical reaction studies for a single chemical system or the study of bulk systems such as water. Bartok et al also suggested alternative representation called smooth overlap of atomic positions (SOAP), where the similarity between two neighborhood environments is directly defined.[28] Very recent work, that introduced a new method known as deep tensor neural networks (DTNN),[15] provides further evidence that NNPs can model a general QM molecular potential when trained to a diverse set of molecular energies. So far, the DTNN model was only trained to small test data sets to show the model could predict molecular energies in specific cases, i.e. equilibrium geometries of organic molecules or the energy along the path of short QM molecular dynamics trajectories. In our experience, training to trajectories can bias the fitness of a model to the specific trajectory used for training, especially along short trajectories. Also, DTNN was not shown to predict energies for larger systems than those included in the training set.

Since the introduction of BP SFs, they have been employed in numerous studies where neural network potentials (NNP) are trained to molecular total energies sampled from MD data to produce a function that can predict total energies of molecular conformations outside of the training set. In general, the NNPs developed in these studies are non-transferable, aside from bulk materials[25,29] and water cases[30]. None of the studies that utilize the SFs of Behler and Parrinello have presented a NNP that is truly transferable between complex chemical environments, such as those found in organic molecules, aside from one limited case of all trans-alkanes[31] where non-equilibrium structures and potential surface smoothness are not considered. We attribute two reasons for the lack of transferability of the SFs. First, as originally defined, SFs lack the functional form to create recognizable features (spatial arrangements of atoms found in common organic molecules, e.g. a benzene ring, alkenes, functional groups) in the molecular representation, a problem that can prevent a neural network from learning interactions in one molecule and then transferring its knowledge to another molecule upon prediction. Second, the SFs have limited atomic number differentiation, which empirically hinders training in complex chemical environments. In general, the combination of these reasons limit the original SFs to studies of either chemically symmetric systems with one or two atom types or very small single molecule data sets.

In this work, we present a transferable deep learning[32,33] potential that is applicable to complex and diverse molecular systems well beyond the training data set. We introduce ANAKIN-ME (Accurate NeurAl networK engINe for Molecular Energies) or ANI in short. ANI is a new method for developing NNPs that utilizes a modified version of the original SFs to build single-atom atomic environment vectors (AEV) as a molecular representation. AEVs solve the transferability problems that hindered the original Behler and Parrinello SFs in complex chemical environments. With AEVs, the next goal of ANI becomes to sample a statistically diverse set of molecular interactions, within a domain of interest, during the training of an ANI class "potential" to produce a transferable NNP. This requires a very large data set that spans molecular conformational and configurational space, simultaneously. An ANI potential trained in this way is well suited to predict energies for molecules within the desired training set domain (organic molecules in this paper), which is shown to be extensible to larger molecules than those included in the training set.

ANI uses an inherently parallel computational algorithm. It is implemented in an in-house software package, called NeuroChem, which takes advantage of the computational power of graphics processing units (GPU) to accelerate the training, testing, and prediction of molecular total energies via an ANI potential. Finally, we show the accuracy of ANI-1 compared to its reference DFT level of theory and, for context, three popular semi-empirical QM methods, AM1, PM6, and DFTB, through four case studies. All case studies only consider larger organic molecules than ANI-1 was trained to predict energies for, providing strong evidence of the transferability of ANI-1.

## 2 Theory and neural network potential design

### 2.1 Neural network potentials

Deep learning[33] is a form of machine learning model that uses a network of computational neurons, which are organized in layers. Specifically, ANI uses a fully-connected *neural network (NN) model in this work.* NNs are highly flexible, non-linear functions with optimizable parameters, called weights, which are updated through the computation of analytic derivatives of a cost function with respect to each weight. The data set used to optimize the weights of a NN is called a training set and consists of inputs and a label, or reference value, for each input. Multi-layered NNs are known as universal function approximators[34] because of their ability to fit to arbitrary functions. A neural network potential[35,36] (NNP) utilizes the regression capabilities of NNs to predict molecular potential surfaces given only information about the structure and composition of a molecule. Standard NNPs suffer from many problems that need to be solved before any generalized model can be built. First, training neural networks to molecules with many degrees of freedom (DOF) is difficult because the data requirements grow with each DOF to obtain a good statistical sampling of the potential energy surface. Also, the typical inputs, such as internal coordinates or coulomb matrices, lack transferability to different molecules since the input size to a neural network must remain constant. Finally, the exchange of two identical atoms in a molecule must lead to the same result.

## 2.2 The ANAKIN-ME model

Heavily modified Behler and Parrinello symmetry functions[25] (BPSF) and their high-dimensional neural network potential model, depicted in Figure 1, form a base for our ANAKIN-ME (ANI) model. The original BPSFs are used to compute an atomic environment vector (AEV), $\vec{G}_i^X = \{G_1, G_2, G_3, \cdots, G_M\}$, composed of elements, $G_m$, which probe specific regions of an individual atom's radial and angular chemical environment. Each $\vec{G}_i^X$ for the $i^{th}$ atom of a molecule with atomic number $X$ is then used as input into a single NNP. The total energy of a molecule, $E_T$, is computed from the outputs, $E_i$, of the atomic number specific NNPs by,

$$E_T = \sum_i^{\text{All Atoms}} E_i \quad (1)$$

In this way, $E_T$ has the form of a sum over all $i$ "atomic contributions" to the total energy. Aside from transferability, an added advantage of this simple summation is that it allows for a near linear scaling in computational complexity with added cores or GPUs, up to the number of atoms in the system of interest.

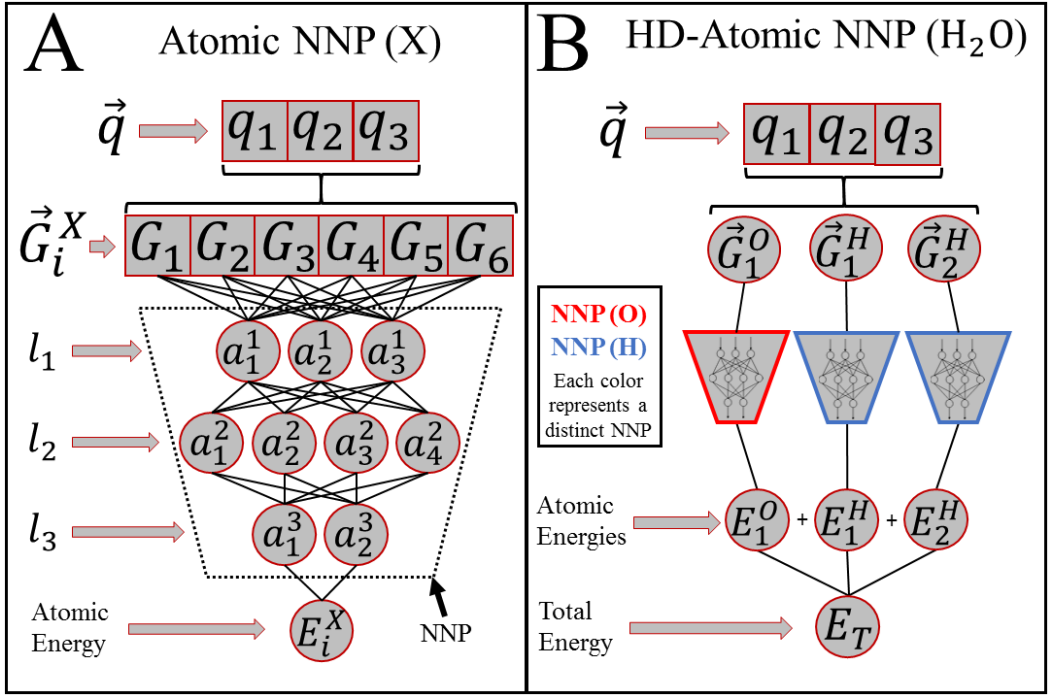

*Figure 1: Behler and Parrinello's HDNN or HD-Atomic NNP model. A) A scheme showing the algorithmic structure of an atomic number specific neural network potential (NNP). The input molecular coordinates, $\vec{q}$, are used to generate the atomic environment vector, $\vec{G}_i^X$, for atom i with atomic number X. $\vec{G}_i^X$ is then fed into a neural network potential (NNP) trained specifically to predict atomic contributions, $E_i^X$, to the total energy, $E_T$. Each $l_k$ represents a hidden layer of the neural network and are composed of nodes denoted by $a_j^k$ where j indexes the node. B) The high-dimensional atomic NNP (HD-Atomic NNP) model for a water molecule. $\vec{G}_i^X$ is computed for each atom in the molecule then input into their respective NNP (X) to produce each atom's $E_i^X$, which are summed to give $E_T$.*

The $\vec{G}_i^X$ vectors are key to allowing this functional form of the total energy to be utilized. For an atom $i$, $\vec{G}_i^X$ is designed to give a numerical representation, accounting for both radial and angular features, of $i$'s local chemical environment. The local atomic environment approximation is achieved with a piece-wise cutoff function,

$$f_C(R_{ij}) = \begin{cases} 0.5 \times \cos\left(\frac{\pi R_{ij}}{R_C}\right) + 0.5 & \text{for } R_{ij} \leq R_C \\ 0.0 & \text{for } R_{ij} > R_C \end{cases} \quad (2)$$

Here, $R_{ij}$ is the distance between atoms $i$ and $j$, while $R_c$ is a cutoff radius. As written, $f_C(R_{ij})$ is a continuous function with continuous first derivatives.

To probe the local radial environment for an atom $i$, the following radial symmetry function, introduced by Behler and Parrinello, produces radial elements, $G_m^R$ of $\vec{G}_i^X$,

$$G_m^R = \sum_{j \neq i}^{\text{All Atoms}} e^{-\eta(R_{ij}-R_s)^2} f_C(R_{ij}) \qquad (3)$$

The index $m$ is over a set of $\eta$ and $R_s$ parameters. The parameter $\eta$ is used to change the width of the Gaussian distribution while the purpose of $R_s$ is to shift the center of the peak. In an ANI potential, only a single $\eta$ is used to produce thin Gaussian peaks and multiple $R_s$ are used to probe outward from the atomic center. The reasoning behind this specific use of parameters is two-fold: first, when probing with many small $\eta$ parameters, vector elements can grow to very large values, which are detrimental to the training of NNPs. Second, using $R_s$ in this manner allows the probing of very specific regions of the radial environment, which helps with transferability. $G_m^R$, for a set of $M = \{m_1, m_2, m_3, ...\} = \{(\eta_1, R_{S_1}), (\eta_2, R_{S_2}), (\eta_3, R_{S_3}), ...\}$ parameters, is plotted in Figure 2A. M consist of a constant $\eta$ for all $m$ and multiple $R_s$ parameters to show a visualization of how each vector element probes its own distinct region of an atom's radial environment.

We made two modifications to the original version of Behler and Parrinello's angular symmetry function to produce one better suited for probing the local angular environment of complex chemical systems. The first addition is $\theta_s$, which allows an arbitrary number of shifts in the angular environment, and the second is a modified exponential factor that allows an $R_S$ parameter to be added. The $R_S$ addition allows the angular environment to be considered within radial shells based on the average of the distance from the neighboring atoms. The effect of these two changes is that AEV elements are generally smaller because they overlap atoms in different angular regions less and they provide a distinctive image of various molecular features, a property that assists neural networks in learning the energetics of specific bonding patterns, ring patterns, functional groups, or other molecular features.

Given atoms $i$, $j$, and $k$, an angle $\theta_{ijk}$ centered on atom $i$ is computed along with two distances $R_{ij}$ and $R_{ik}$. A single element, $G_m^{A\text{mod}}$ of $\vec{G}_i^X$, to probe the angular environment of atom $i$ takes the form of a sum, over all j and k neighboring atom pairs, of the product of a radial and an angular factor,

$$G_m^{A\text{mod}} = 2^{1-\zeta} \sum_{j,k \neq i}^{\text{All Atoms}} (1 + \cos(\theta_{ijk} - \theta_s))^\zeta \exp\left[-\eta\left(\frac{R_{ij}+R_{ik}}{2} - R_S\right)^2\right] f_C(R_{ij}) f_C(R_{ik}) \qquad (4)$$

The Gaussian factor combined with the cutoff functions, like the radial symmetry functions, allows chemical locality to be exploited in the angular symmetry functions. In this case, the index $m$ is over four separate parameters: $\zeta$, $\theta_s$, $\eta$, and $R_S$. $\eta$ and $R_S$ serve a similar purpose as in Equation 3. Applying a $\theta_s$ parameter allows probing of specific regions of the angular environment in a similar way as is accomplished with $R_S$ in the radial part. Also, $\zeta$ changes the width of the peaks in the angular environment. $G_m^{A\text{mod}}$ for several $m$ are plotted in Figure 2B while the original angular function is plotted in Figure 2C. With the original Behler and Parrinello angular function, only two shifting values were possible in the angular environment, 0 and $\pi$. The modified angular function allows an arbitrary number to be chosen, allowing for better resolution of the angular environment. As with its radial analog, this helps to keep the elements of $\vec{G}_i^X$ small for better NNP performance and allows probing of specific regions of the angular chemical environment.

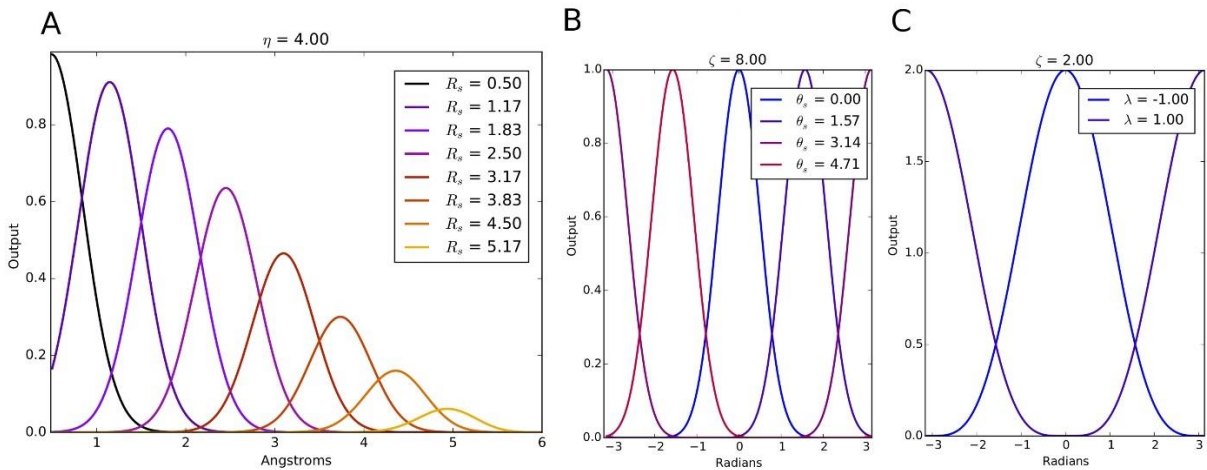

*Figure 2: Examples of the symmetry functions with different parameter sets. A) Radial symmetry functions, B) Modified angular symmetry functions and C) the original Behler and Parrinello angular symmetry functions. These figures all depict the use of multiple shifting parameters for each function, while keeping the other parameters constant.*

### 2.2.1 Atomic number differentiated atomic environment vector

In this work, we differentiate between atomic numbers in the AEV by supplying a radial part for each atomic number and an angular part for each atomic number pair in the local chemical environment. The original BPSFs treat all atoms identically in the summation over all atoms, and thus individual atomic number specific NNPs are unable to distinguish between a carbon, hydrogen, or any other atom type at some distance. Through empirical evidence, provided in Table S4 of the supplemental information (SI), we have found that discriminating between atomic numbers allows for training to much lower error on diverse multi-molecule training sets and permits better transferability.

For AEVs built from $N$ atom types, this leads to $N$ radial sub-AEVs and $N(N + 1)/2$ angular sub-AEVs. SI Figure S1 gives an example of an atomic number differentiated AEV for the carbon atom in formic acid with only 8 radial symmetry functions and 8 angular symmetry functions. The figure shows an overlay of two AEVs each representing a different C-O-H angle with the rest of the structure frozen. From this figure, it is easy to identify the different features which represent formic acid and it also provides clear information on the conformation of the molecule. It is this clearly defined "fingerprint" that allows the modified symmetry functions to perform well in such diverse chemical environments.

## 2.3 Normal mode sampling

The ANI method requires many training and testing data points, $(\vec{q}, E_T)$, where $\vec{q}$ is some energy minimized or non-minimized molecular coordinates, a conformation, from a diverse set of molecules and $E_T$ is the single point energy calculated at a desired QM level of theory. To obtain an accelerated but physically relevant sampling of molecular potential surfaces, we propose the Normal Mode Sampling (NMS) method to generate structures for which single point energies can be computed. A method akin to our version of normal mode sampling has successfully been employed in generating non-equilibrium structures in order to obtain a data set of atomic forces for training a ML model.[21] The end goal of NMS is to generate a set of data points on the potential surface, or a window, around a minima energy structure of a molecule out to some maximum energy. Using the proposed NMS gives some confidence that interactions to a specific temperature are accounted for in a trained ANI potential.

To carry out normal mode sampling on an energy minimized molecule of $N_a$ atoms, first a set of $N_f$ normal mode coordinates, $Q = \{q_1, q_2, q_3, \dots q_{N_f}\}$, is computed at the desired *ab-initio* level of theory, where $N_f = 3N_a - 5$ for linear molecules and $N_f = 3N_a - 6$ for all others. The corresponding force constants, $K = \{K_1, K_2, K_3, \cdots, K_{N_f}\}$, are obtained alongside $Q$. Then a set of $N_f$ uniformly distributed pseudo-random numbers, $c_i$, are generated such that $\sum_i^{N_f} c_i$ is in the range [0,1]. Next, a displacement, $R_i$, for each normal mode coordinate is computed by setting a harmonic potential equal to the $c_i$ scaled average energy of the system of particles at some temperature, T. Solving for the displacement gives,

$$R_i = \pm \sqrt{\frac{3c_i N_a k_b T}{K_i}} \quad (5)$$

where $k_b$ is Boltzmann's constant. The sign of $R_i$ is determined randomly from a Bernoulli distribution where $p = 0.5$ to ensure that both sides of the harmonic potential are sampled equally. The displacement is then used to scale each normalized normal mode coordinate by $q_i^R = R_i q_i$. Next, a new conformation of the molecule is generated by displacing the energy minimized coordinates by $Q^R$, the superposition of all $q_i^R$. Finally, a single point energy at the desired level of theory is calculated using the newly displaced coordinates as input.

The choice of temperature is dependent on the intended use of the ANI potential being trained. However, it should be noted that this method of sampling the potential surface of a molecule is simply an approximation for generating structures. In practice, NMS works best when generating windows of the potential surface of many molecules to be used in the training of the same ANI potential. The reasoning behind this is as follows: if any interactions are missed or not sampled well by NMS, it is possible that other molecules in the data set contain the same or similar interactions. Therefore, the accuracy of using such a sampling method is dependent on not only the number of points per window but also the number of distinct molecules included in the data set.

## 3 Methods

### 3.1 Data Selection

The accuracy of any empirical potential, especially an ANI potential, is highly dependent on the amount, quality of, and types of interactions included in the data used to train the model. For instance, a data set generated from high level CCSD(T) *ab-initio* theory, for every possible combination of all atom types and a full sampling of configurations in three-dimensional space would be ideal for training an ANI potential. However, this is not possible due to time and other practicality considerations. Therefore, we limit the scope of this study to a specific class of systems, namely organic molecules with four atom types: H, C, N, and O. We also restrict our data set to near equilibrium conformations since a full sampling of each structure's potential surface increases the number of data points required for training to a near intractable level. Data sets have been developed[37] with a similar search of chemical space, however, these data sets only cover configurational space and not conformational space, which is a requirement for training an ANI class potential. In this work, we choose ωB97X[38], the hybrid meta-GGA DFT functional, with the 6-31G(d) basis set as reference QM data. The ωB97X functional provides excellent accuracy for molecular structures, stability, bond energies and reaction barriers. Everything described in this article can be repeated at any other level of QM theory if wanted.

### 3.2 The GDB-11 database

A good starting point to build a training data set for organic molecules is the GDB-11 database[39,40]. The GDB-11 database is built from all possible molecules containing up to 11 atoms of the atomic numbers C, N, O, and F and is filtered by chemical stability and synthetic feasibility considerations, as well as simple valency rules. Molecules in GDB-11 are supplied in the form of SMILES strings[41], which we converted to 3D structures using the RDKit software package[42].

The ANI-1 data set employed in this work, was generated from a subset of the GDB-11 database containing molecules without the fluorine atom. This leaves only molecules with H, C, N, and O after hydrogens are added with RDKit. Also, given the sheer number of molecules (40.3 million) in the GDB-11 database, as of the time of this article, only reference data for molecules up to 8 atoms of C, N, and O have been computed. In total, 57,951 molecules are included in our current data set, the ANI-1 data set. A breakdown of how many molecules are included from each GDB-11 subset is given in SI Table S1. All energies are computed with neutral molecules in the singlet spin state.

### 3.3 ANI-1 data set generation

From a proper database of molecules within a chemical domain of interest, a data set must be generated that includes sampling of each molecule's potential surface around its equilibrium structure. We do this in the spirit of work carried out by many others[16,35,36,43] who fitted neural networks to single molecule potential surfaces. Given simple physical considerations, the sampling of the potential surface can be limited to a window of relevant energies. Sampling can be carried out using quantum mechanical (QM) molecular dynamics (MD) simulation as suggested by others.[44] However, QM MD is inefficient for producing a small data set from a sampling of a large window of a potential surface, which is desirable for the ANI method. The reason for this is that configurationally diverse data sets overlap interactions throughout the data set, so larger molecules require far less data points (~200) than smaller ones. Because of this, utilizing MD would follow a well-defined trajectory along the potential surface and would lead to sampling biased to the specific trajectory. Thus, a very long trajectory is required to overcome this bias. It is for this reason that a more stochastic natured sampling is required for the ANI method.

In this work, we propose a Normal Mode Sampling (NMS) method that works by calculating the normal modes of a molecule, then randomly perturbing the equilibrium structure along these normal modes out to a maximum energy (see section 2.3 for details on NMS). The ANI-1 data set was generated by applying NMS to every molecule with 8 or less heavy atoms in the GDB-11 database. Using the wB97X[38] DFT functional with the 6-31G(d) basis set in the Gaussian 09 electronic structure package[45], the following steps are followed to generate the data set:

1. Convert SMILES strings to 3D structures and add hydrogens to fill valence orbitals.
2. Optimize each molecule in the database using tight convergence criteria.
3. Generate normal modes for each optimized molecule with an ultra-fine DFT grid.
4. Use NMS method to generate K structures for each molecule in the database. The exact number of structures per molecule is determined by $K = S(3N - 6)$. S is an empirically determined value dependent on the number of heavy atoms in the molecule and N is the total number of atoms in the molecule, including hydrogens.
5. Calculate single point energies for each of the generated structures.

Using this procedure to generate the ANI-1 data set results in molecular energies for a total of ~17.2 million conformations generated from ~58k small molecules. For each molecule's individual set of random conformations, 80% is used for training, while 10% is used for each validation and testing of the ANI-1 model.

For practical considerations, the value S from step 3 is large (about 500) for very small molecules and is gradually reduced as the number of heavy atoms, and molecule diversity, grows. Table S1 in the SI shows the parameters used in the production of the ANI-1 data set, including the S values used for each GDB-11 database subset as well as the per atom test set RMSE of an ANI potential vs DFT for each subset.

### 3.4 Training the ANI-1 potential

All ANI potential training, validating, and predicting is done with an in-house C/C++ and CUDA GPU accelerated software package we call NeuroChem (C++ interface) and pyNeuroChem (Python interface). Where applicable, the neural network algorithm is encoded as either matrix-matrix, matrix-vector, or vector-vector operations using CUBLAS[46]. The atomic environment vectors are computed through a separate in-house built library called AEVLib, which is also GPU accelerated.

Finding a good set of atomic environment vector ($\vec{G}$) parameters to compute molecular representations plays a major role in how well the ANI-1 potential trains and performs. Too many $\vec{G}$ parameters will lead to networks that are very large, and thus hard to train. Too few parameters result in low resolution of the local chemical environment, which is detrimental to transferability and training in general. For the ANI-1 potential, 32 evenly spaced radial shifting parameters are used for the radial part of $\vec{G}$ and a total of 8 radial and 8 angular shifting parameters are used for the angular part. The specific AEV parameters were chosen with a few goals in mind: to minimize the size of the AEV, to maximize the resolution of the local atomic environments, and to cover all space within the cutoff radius provided. Keeping these goals in mind the choice of parameters can be automated to simply chose multiple $R_s$ and $\theta_s$ parameters equally spaced and setting the η and ζ parameters such that one function overlaps with its neighboring function slightly, as shown in Figure 2. With four atom types, this leads to a total of 768 elements in $\vec{G}$. The cutoff radii of 4.6Å for the radial and 3.1Å for the angular symmetry functions were chosen based on the distribution of atomic distances and an assumption that angular environments are less sampled in the ANI-1 data set, empirical testing verified this to be the case.

The choice of network architecture also plays a major role in how well a potential performs. Too small of a network reduces the flexibility of the function which can hinder performance and too large can lead to bad generalization across structures due to overtraining, especially on small data sets. With larger data sets, a bigger and more flexible network can be used to yield better results. We empirically tested many network architectures. Generally, 3 to 4 hidden layer networks with between 32 and 128 nodes per layer performed the best. The best ANI model (ANI-1), employed in this work, was trained to 80% of the 17+ M data points, and has the following pyramidal architecture: 768:128:128:64:1. That is, 768 input values followed by a 128-node hidden layer followed by another hidden layer with 128 nodes, a 64-node hidden layer, and finally a single output node for a total of 124,033 optimizable parameters per each individual atomic number neural network potential. All hidden layer nodes use a Gaussian activation function[47] while the output node uses a linear activation function. The weights are randomly initialized from a normal distribution in the range $(-1/\sqrt{d}, 1/\sqrt{d})$, where d is the number of inputs into the node. The neural network bias parameters are all initialized to zero.

To train the weights, the program randomly samples structures from the training set in a mini-batch of 1024 molecules. Next a cost derivative w.r.t. each weight is calculated through back-propagation from the exponential cost function[48],

$$C(\vec{E}^{ANI}) = \tau \exp\left(\frac{1}{\tau}\sum_j (E_j^{ANI} - E_j^{DFT})^2\right) \qquad (6)$$

$\vec{E}^{ANI}$ is a vector of the energy outputs, $E_j^{ANI}$, from the ANI network for the $j^{th}$ set of coordinates. $E_j^{DFT}$ are the corresponding DFT reference energies. The parameter $\tau$ is set to 0.5 for best performance. This cost function was chosen because of its robustness in handling outliers in data sets, a property that achieves 2 to 4 times lower error upon training an ANI potential. The network weights are optimized via the ADAM update method.[49] An initial learning rate of 0.001 is used with the other ADAM parameters set to $\beta_1 = 0.9$, $\beta_2 = 0.999$, and $\varepsilon = 1.0 \times 10^{-8}$, as recommended by the ADAM authors. To avoid node saturation the incoming weight vector to each node in the network is constrained by the max norm regularization method[50] to a maximum length of 3.0. The mini-batch update is repeated over the full training set until a training epoch is completed. Training epochs are iterated until the validation set stops improving in accuracy for 100 epochs. The optimization process is carried out 6 times using an order of magnitude smaller learning rate each time. The final fitness of the training, validation, and test sets in the case of the ANI-1 potential are 1.2, 1.3, and 1.3 root mean squared error (RMSE) in kcal/mol, respectively.

## 4 Results and discussion

The final ANI potential for the domain of organic molecules containing the atoms H, C, N, and O, is trained over a data set containing over 80% of the 17.2 million data points in the ANI-1 data set. This data set, produced by applying normal mode sampling (NMS, developed in the present work) to more than 56k distinct small molecules from the GDB-8 database, spans the configurational as well as conformational space of organic molecules. Such vast data is required to ensure the sampling of relevant interactions needed to produce a very high dimensional potential surface. Figure 3 stands as evidence to the necessity of this vast amount of training data. More important than the low errors to the training, validation, and test sets, it shows that the extensibility of ANI potentials increase with data set size, and does not seem to plateau even up to the current data set size.

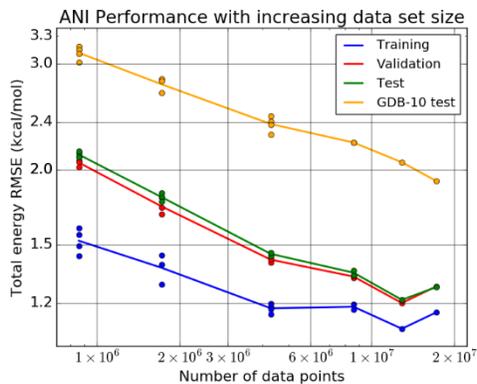

*Figure 3: Log-log plots of the training, validation, testing, and a random GDB-10 (molecules with 10 heavy atoms from the GDB-11 database) extensibility testing set total energy errors vs. increasing number of data points in the training set. The sets of points converge to the final ANI-1 potential presented in this paper, trained on the full ANI-1 data set.*

We performed extensive benchmark and case studies to estimate the accuracy of the ANI-1 potential compared to DFT reference calculations. As baselines, in the first test case we compare ANI-1 to a sorted coulomb matrix[13] (CM) molecular representation with a multilayer perceptron (MLP) neural network model, baseline 1, and to an ANI type neural network model trained where the AEVs are not type differentiated, baseline 2. MLP's were chosen in baseline 1 because of their ability to train to very large data sets via batched learning. Table S4 in the SI provides details of these baselines for comparison to the ANI method.

To highlight the true transferability of the ANI-1 potential, all molecules considered in the following test cases are *greater than eight heavy atoms*. The atom counts for these test systems range from 10 to 24 heavy atoms up to a total of 53 atoms. First, we analyzed ANI-1's overall performance, goodness of fit, and transferability to non-minimized structures with a total of 8245 conformations generated using NMS on 134 randomly selected molecules from GDB-11, each with 10 heavy atoms. In the second case study, we look at the accuracy of ANI-1 in predicting the relative energies of DFT energy minimized $C_{10}H_{20}$ isomers with respect to the lowest energy isomer. Third, energy differences are compared for energy minimized conformers of the drug molecule Retinol. And finally, four rigid scans, a bond stretch, an angle bend, and two dihedral rotations on relatively large drug molecules are carried out on ANI-1 and compared with reference DFT results. For comparison, we also show performance of popular DFTB, PM6, and AM1 semi-empirical methods in all the test cases presented.

## 4.1 Statistical fitness

To show the overall accuracy and transferability of the ANI-1 potential, Figure 4 plots the energy correlation of relative energies for a subset of molecules from the GDB-11 database. Specifically, the sampling includes 8,245 total NMS generated conformations and their respective energies from 134 randomly selected molecules with 10 heavy atoms. This gives a set of 62 conformations, on average, per molecule. Each molecule's test set is shifted such that the lowest energy is zero to compare relative energies. An absolute energy comparison of this test set, between ANI-1 and DFT, is provided in SI table S2.

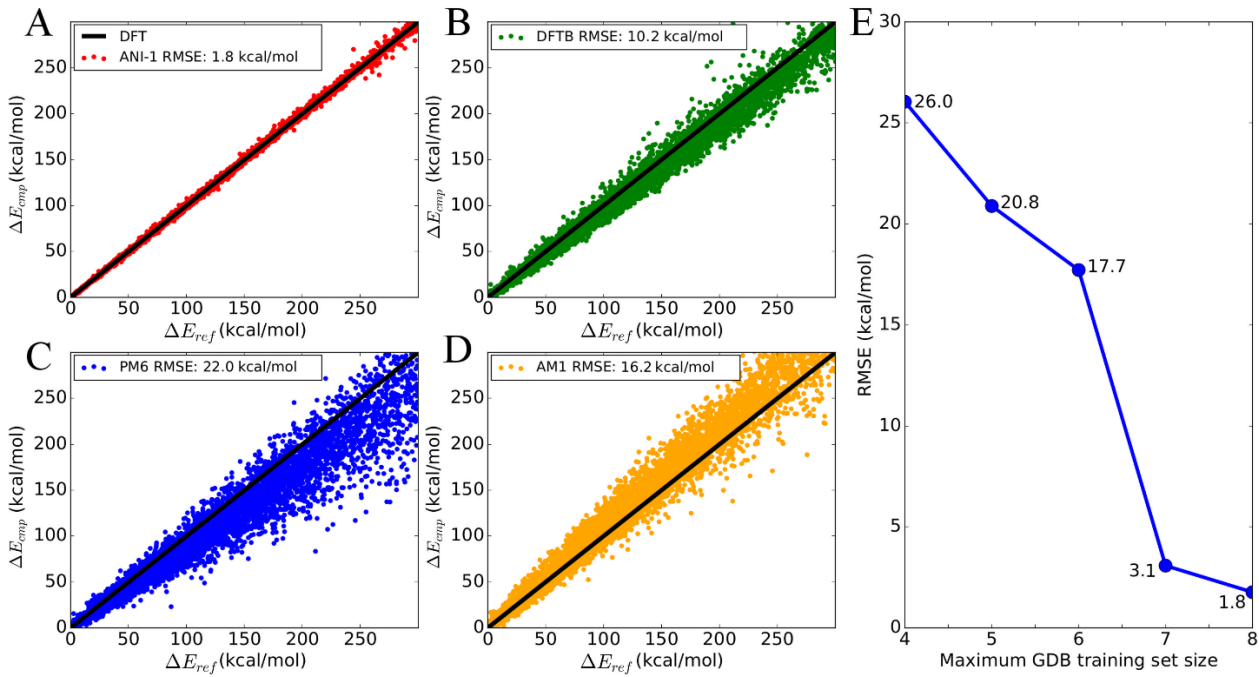

*Figure 4: Relative energy comparisons from random conformations of a random sampling of 134 molecules from GDB-11 all with 10 heavy atoms. There is an average of 62 conformations, and therefore energies, per molecule. Each set of energies for each molecule is shifted such that the lowest energy is at 0. None of the molecules from this set are included in any of the ANI training sets. A-D) Correlation plots between DFT energies, $E_{ref}$, and computed energies, $E_{cmp}$, for ANI-1 and popular semi-empirical QM methods. Each individual molecule's set of energies is shifted such that the lowest energy is at zero. E) RMS error (kcal/mol) of various ANI potentials, compared to DFT, trained to an increasing data set size. The x-axis represents the maximum size of GDB molecules included in the training set. For example, 4 represents an ANI potential trained to a data set built from the subset of GDB-11 containing all molecules up to 4 heavy atoms.*

Figure 4A is a correlation plot of the computed ANI-1 energies, $E_{cmp}$, vs. the DFT reference energies, $E_{ref}$. The ANI-1 potential achieves an RMSE of only 1.8 kcal/mol over the entire random sampling Figure 3B-D provides the same comparison but for popular semi-empirical methods to the DFT reference energies. If only relative energies within 30 kcal/mol of the minimum energy are considered, the ANI-1, DFTB, PM6, and AM1 methods obtain an RMSE of 0.6, 2.4, 3.6, and 4.2 kcal/mol, respectively. SI table S3 lists the total energy and relative energy error of the ANI-1 potential as an energy cap, $E^{cap}$, is lowered until finally only minimum energy structures are considered.

Figure 4E shows how the RMSE of an ANI potential to reference DFT decreases as the number of distinct molecules grow in the training set. From this plot, it is clear that the addition of more data leads to better fits, with the largest and most diverse data set achieving an RMSE of just 1.8 kcal/mol. Inclusion of molecules with 7 heavy atoms, mostly mono-substituted aromatic compounds, yields a dramatic reduction of the RMSE. This figure, along with Figure 3, stands as evidence that increasing the size and diversity of an ANI training set leads to better fitness and transferability, meaning future parametrization will yield even better results.

The total energies produced by ANI-1, baseline 1, and baseline 2 for the GDB-10 test set are also compared. ANI-1, when trained on the full ANI-1 training set, achieves a total energy RMSE of 1.9 kcal/mol while baseline 1 and baseline 2 achieve a RMSE of 493.7 kcal/mol and 6.7 kcal/mol, respectively. While the baselines perform better on the ANI-1 test set, as seen in SI Figure S4, the data above shows that both suffer from an inability to extend their learned interactions to larger molecules. For baseline 1, this is caused by the coulomb matrix having elements which remain a constant zero throughout training, yet when a larger molecule is tested on it, those elements have non-zero values. These non-zero values are then fed into untrained network parameters, which yields arbitrary results. For baseline 2, the problem comes from the fact that the AEVs have an inability to differentiate between atom types, creating confusion during the learning process.

### 4.2 Structural and geometric isomers

This case study looks at relative stabilities of structural and geometric isomers with the empirical formula $C_{10}H_{20}$. All isomers were optimized at the chosen DFT level of theory. Structures of all isomers included in this case study are shown in SI Figure S2. Figure 5 gives a visual comparison of the ANI-1 potential and different semi-empirical methods to DFT calculated energies of the isomers. The energies are ordered from the lowest to the highest for clarity. The x-axis shows the isomer index number, which matches to the molecule index in SI Figure S2.

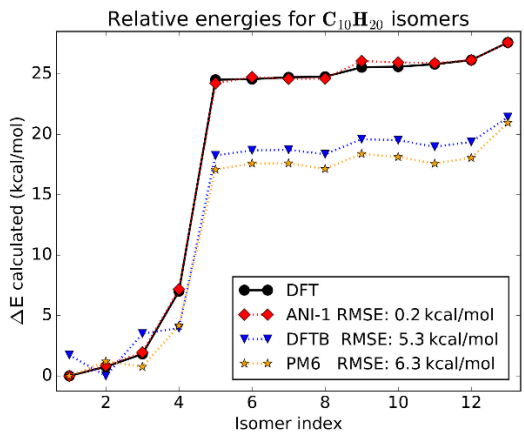

*Figure 5: The total energies, shifted such that the lowest is zero, calculated for various $C_{10}H_{20}$ isomers are compared between DFT with the ωB97X functional and 6-31g(d) basis set, the ANI-1 potential, AM1 semi-empirical, and PM6 semi-empirical methods.*

Figure 5 shows that the ANI-1 potential properly predicts the minimum energy structure and continues to match the energies of the ring containing structures, indices 1–4 on the x-axis, with a very low error and with proper ordering. Also, when moving from the ringed structures to the linear alkenes, index 4 to 5, the ANI-1 potential approximates the DFT energy difference between these two classes of molecules very well. The linear alkanes, indices 5 – 13, fit very well to the DFT energies. Overall the ANI-1 potential achieves an RMSE of 0.2 kcal/mol. In contrast, both DFTB and PM6 methods incorrectly predict relative stability of ring containing structures. Energies of isomers 5-13 are systematically underestimated by about 6-7 kcal/mol.

### 4.3 Conformers of Retinol

Eight conformers of the molecule Retinol were generated using the RDKit package and then optimized to their respective DFT energy minima. In this case study, Figure 6, the energy difference, $\Delta E$, and $|\Delta\Delta E|$ are plotted to show how well the ANI-1 potential performs at predicting energy differences when large conformational changes, i.e. many dihedral rotations over the entire molecule occur. The $|\Delta\Delta E|$ plots represent the absolute value of the differences between the elements of the DFT plot and the elements of the other method's $\Delta E$ plots. All $\Delta E$ plots are on the same scale, shown to the right of the figures, and the same is true for the $|\Delta\Delta E|$ plots.

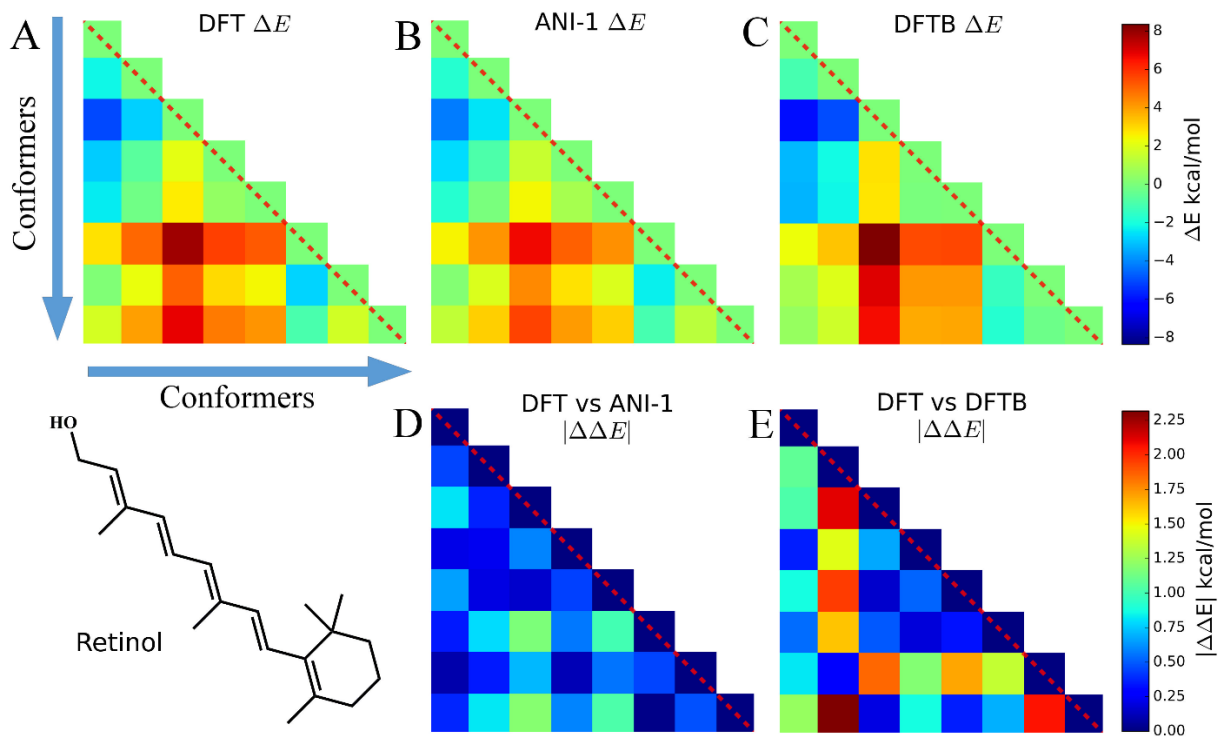

*Figure 6: A-C) These three triangle plots, which are on the same scale shown to the right, show energy differences, ΔE, between random energy minimized conformers of the molecule retinol. The structural differences between these conformers include many dihedral rotations. A) shows the conformers ΔE calculated with DFT, B) ANI-1, and C) DFTB. D shows the absolute value of the difference between A and B, |ΔΔE|, while E shows the same between A and C. ΔE and |ΔΔE| have their own scale shown to the right of the plots. All plots of a specific type use the same color scaling for easy comparison.*

Figure 6A shows ΔE between each Retinol conformer for DFT while B shows the same for ANI-1 and C for DFTB. Aside from some minor shading differences comparing A and B clearly shows how well the ANI-1 energy differences match that of the DFT calculations. Figure 6D and E contain |ΔΔE| plots corresponding to A vs. B and A vs. C, respectively, and shows that the ANI-1 potential can predict DFT energy differences of these large structural changes to a very low error. In total, ANI-1 and DFTB achieve a RMSE to the DFT ΔE of 0.6 kcal/mol and 1.2 kcal/mol, respectively. However, DFTB severely over estimates energies of conformers 2 and 7.

### 4.4 Potential surface accuracy

So far, all test cases have only considered large structural changes or unordered NMS generated structures. However, to be useful in molecular dynamics simulations, the ANI-1 potential must not only have a low error, but must also produce a very smooth physically meaningful surface. To provide evidence that ANI-1 satisfies these requirements, unrelaxed scans were conducted on different drug molecules and are plotted in Figure 7.

Figure 7A shows a bond stretch, from 1.15Å to 1.75Å, of the N-C bond (labeled 1 and 2) in the analgesic drug molecule Fentanyl[51]. The bond equilibrium distance was calculated separately for each method and was found to be 1.38Å for DFT, 1.39Å for ANI-1, 1.40Å for DFTB, and 1.41Å for PM6. Figure 7 presents an angle bend, from 90.0° to 135.0°, for the C-C-C angle labeled 1-2-3 in the structure of Fentanyl included within the plot. As with the bond stretch, the ANI-1 potential produces an angle bend potential surface with a very low RMSE of only 0.4kcal/mol while maintaining a very smooth curvature for accurate force calculations. ANI-1 produces an angle bend potential with an equilibrium angle ~1.0° from the DFT equilibrium. PM6 and DFTB produce equilibrium structures at ~1.05° and ~0.75°, respectively, from the DFT calculation.

Finally, Figure 7C and D depict rotations of the dihedral angles labeled in the two figures. Figure 7C shows a C-C-C-C dihedral rotation potential in the molecule 4-Cyclohexyl-1-butanol, while Figure 7D is for an N-C-C-C dihedral angle in the drug molecule called Lisdexamfetamine[52]. The ANI-1 potential manages to capture all minima to within 3.0° of the DFT potentials for both plots, which is better or comparable to the semi-empirical methods. As expected both semi-empirical methods severely underestimate dihedral rotation barriers, and in the case of Lisdexamfetamine give an unrealistic shape of potential surface.

Again, both figures not only fit well to the potential surface but model it very well by reproducing the shape and smoothness of the surface. This fact shows that the ANI-1 potential does produce a smooth potential, one that could provide forces, for use in molecular dynamics simulations or optimization problems.

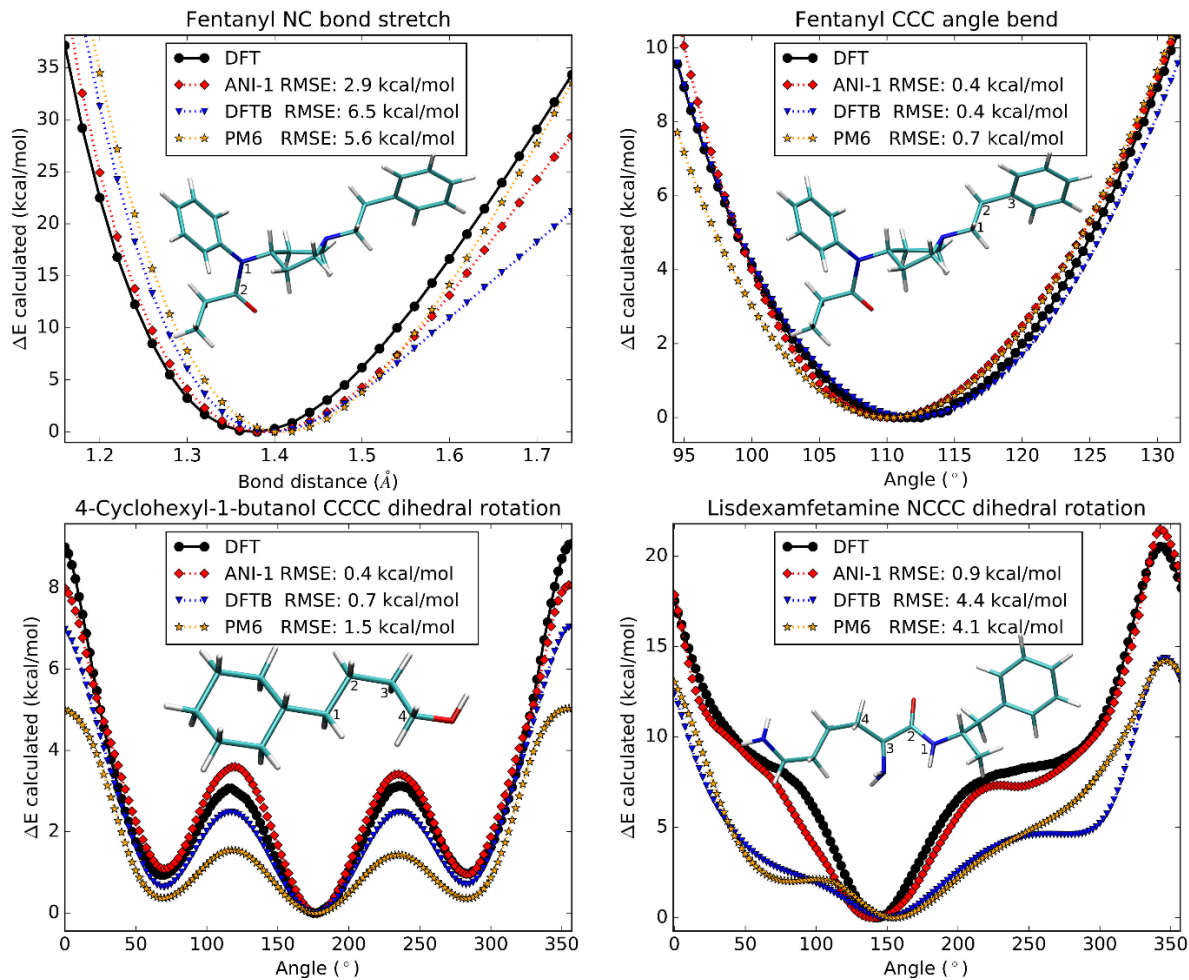

*Figure 7: Each subplot shows a one-dimensional potential surface scan generated from DFT, the ANI-1 potential, and two popular semi-empirical methods, DFTB and PM6. The atoms used to produce the scan coordinate are labeled in the images of the molecules in every sub-plot. Each figure also lists the RMSE, in the legend, for each method compared to the DFT potential surface.*

## 5 Conclusions

In this work we present the first truly transferable neural network potential (NNP) for organic molecules based on a deep learning architecture and with heavy modifications to the HDNN method of Behler and Parrinello[25]. Our NNP, presented as the ANI-1 potential, was trained on a data set, which spans conformational and configurational space, built from small organic molecules of up to 8 heavy atoms. We show its applicability to much larger systems of 10-24 heavy atoms including well known drug molecules and a random selection of 134 molecules from the GDB-11 database containing 10 heavy atoms. ANI-1 shows exceptional predictive power on the 10-heavy atom test set, with RMSE versus DFT relative energies as low as 0.6 kcal/mol when only considering molecular conformations that are within 30 kcal/mol of the energy minimum for each molecule. While the ANI-1 potential specifically targets organic molecules with the atoms H, C, N, and O, the ANI method can be used to build potentials for other classes of molecules and even crystals. ANI-1 was specifically trained to DFT energies, but could be extended to high level *ab-initio* QM methods and larger basis sets given enough computational resources.

As the results clearly show, the ANI method is a potential game changer for molecular simulation. Even the current version, ANI-1, is more accurate vs. the reference DFT level of theory in the provided test cases than DFTB, and PM6, two of the most widely used semi-empirical QM methods. Besides being accurate, a single point energy, and eventually forces, can be calculated in as many as six orders of magnitude faster than DFT. Empirical evidence shows the computational scaling per atom of the method is roughly equivalent to a classical force field for very large molecules.

The accuracy of the ANI method is entirely dependent on the data used during training. Thus, continuing to augment the ANI-1 data set with new molecules and including more atomic numbers will improve the accuracy of the trained ANI potential further as well as extend the method to new chemical environments.

# 6 Acknowledgements

J.S.S. acknowledges the University of Florida for funding through the Graduate School Fellowship (GSF). A.E.R. thanks NIH award GM110077. O.I. acknowledges support from DOD-ONR (N00014-16-1-2311) and Eshelman Institute for Innovation award. Part of this research was performed while O.I. was visiting the Institute for Pure and Applied Mathematics (IPAM), which is supported by the National Science Foundation (NSF). Authors acknowledge Extreme Science and Engineering Discovery Environment (XSEDE) award DMR110088, which is supported by National Science Foundation grant number ACI-1053575. We gratefully acknowledge the support of the U.S. Department of Energy through the LANL/LDRD Program for this work.

# 7 Bibliography

1    K. Kitaura, E. Ikeo, T. Asada, T. Nakano and M. Uebayasi, *Chem. Phys. Lett.*, 1999, **313**, 701–706.

2    D. G. Fedorov, T. Nagata and K. Kitaura, *Phys. Chem. Chem. Phys.*, 2012, **14**, 7562.

3    C. Ochsenfeld, J. Kussmann and D. S. Lambrecht, in *Reviews in Computational Chemistry*, John Wiley & Sons, Inc., 2007, pp. 1–82.

4    M. Elstner, *Theor. Chem. Acc.*, 2006, **116**, 316–325.

5    J. J. P. Stewart, *J. Mol. Model.*, 2009, **15**, 765–805.

6    Dewar M J S, *J. Amer. Chem. Soc.*, 1985, **107**, 3902.

7    W. Thiel, *Perspectives on Semiempirical Molecular Orbital Theory*, John Wiley & Sons, Inc., 2007.

8    T. A. Halgren, *J. Comput. Chem.*, 1996, **17**, 490–519.

9    H. Sun, *J. Phys. Chem. B*, 1998, **5647**, 7338–7364.

10  V. Hornak, R. Abel, A. Okur, B. Strockbine, A. Roitberg and C. Simmerling, *Proteins Struct. Funct. Genet.*, 2006, 65, 712–725.

11  J. A. Maier, C. Martinez, K. Kasavajhala, L. Wickstrom, K. E. Hauser and C. Simmerling, *J. Chem. Theory Comput.*, 2015, **11**, 3696–3713.

12  A. C. T. van Duin, S. Dasgupta, F. Lorant and G. W. A., *J. Phys. Chem. A*, 2001, **105**, 9396–9409.

13  M. Rupp, A. Tkatchenko, K.-R. Muller and O. A. von Lilienfeld, *Phys. Rev. Lett.*, 2012, **108**, 58301.

14  S. Manzhos and T. Carrington, *J. Chem. Phys.*, 2006, **125**, 84109.

15  K. T. Schütt, F. Arbabzadah, S. Chmiela, K. R. Müller and A. Tkatchenko, *arXiv.org*, 2016, 1609.08259.

16  T. H. Ho, N.-N. Pham-Tran, Y. Kawazoe and H. M. Le, *J. Phys. Chem. A*, 2016, **120**, 346–355.

17  B. Kolb, B. Zhao, J. Li, B. Jiang and H. Guo, *J. Chem. Phys.*, 2016, **144**, 224103.

18  B. Jiang, J. Li and H. Guo, *Int. Rev. Phys. Chem.*, 2016, **35**, 479–506.

19  F. Häse, S. Valleau, E. Pyzer-Knapp and A. Aspuru-Guzik, *Chem. Sci.*, 2016, **7**, 5139–5147.

20  F. A. Faber, A. Lindmaa, O. A. von Lilienfeld and R. Armiento, *Phys. Rev. Lett.*, 2016, **117**, 135502.

21  M. Rupp, R. Ramakrishnan and O. A. von Lilienfeld, *arXiv:1505.00350*, 2015, **6**, 1–5.

22  O. Isayev, C. Oses, S. Curtarolo and A. Tropsha, 2016, 1–12.

23  J. Behler, *Int. J. Quantum Chem.*, 2015, **115**, 1032–1050.

24  D. Jasrasaria, E. O. Pyzer-Knapp, D. Rappoport and A. Aspuru-Guzik, 2016.

25  J. Behler and M. Parrinello, *Phys. Rev. Lett.*, 2007, **98**, 146401.

26  O. A. Von Lilienfeld, R. Ramakrishnan, M. Rupp and A. Knoll, *Int. J. Quantum Chem.*, 2015, **115**, 1084–1093.

27  S. Manzhos, R. Dawes and T. Carrington, *Int. J. Quantum Chem.*, 2015, 115, 1012–1020.

28  A. P. Bartók, R. Kondor and G. Csányi, *Phys. Rev. B - Condens. Matter Mater. Phys.*, 2013, **87**, 184115.

29  W. J. Szlachta, A. P. Bartók and G. Csányi, *Phys. Rev. B - Condens. Matter Mater. Phys.*, 2014, **90**, 104108.

30  T. Morawietz, A. Singraber, C. Dellago and J. Behler, *Proc. Natl. Acad. Sci.*, 2016, **113**, 8368–8373.

31  M. Gastegger, C. Kauffmann, J. Behler and P. Marquetand, *J. Chem. Phys.*, 2016, **144**, 194110.

32  Y. LeCun, Y. Bengio, G. Hinton, L. Y., B. Y. and H. G., *Nature*, 2015, **521**, 436–444.

33  J. Schmidhuber, *Neural Networks*, 2015, 61, 85–117.

34  K. Hornik, M. Stinchcombe and H. White, *Neural Networks*, 1989, **2**, 359–366.

35  H. Gassner, M. Probst, A. Lauenstein and K. Hermansson, *J. Phys.*, 1998, **102**, 4596–4605.

36   C. M. Handley and P. L. A. Popelier, *J. Phys. Chem. A*, 2010, **114**, 3371–3383.

37   R. Ramakrishnan, P. O. Dral, M. Rupp and O. A. von Lilienfeld, *Sci. data*, 2014, **1**, 140022.

38   J. Da Chai and M. Head-Gordon, *J. Chem. Phys.*, 2008, **128**, 84106.

39   T. Fink and J. L. Raymond, *J. Chem. Inf. Model.*, 2007, **47**, 342–353.

40   T. Fink, H. Bruggesser and J. L. Reymond, *Angew. Chemie - Int. Ed.*, 2005, **44**, 1504–1508.

41   *www.opensmiles.org*.

42   G. Landrum, *www.rdkit.org*.

43   J. Behler, *Phys. Chem. Chem. Phys.*, 2011, **13**, 17930.

44   L. M. Raff, M. Malshe, M. Hagan, D. I. Doughan, M. G. Rockley and R. Komanduri, *J. Chem. Phys.*, 2005, **122**, 84104.

45   G. M. J. Frisch, W. Trucks, H. B. Schlegel, G. E. Scuseria, M. A. Robb, J. R. Cheeseman, G. Scalmani, V. Barone, B. Mennucci, G. A. Petersson, H. Nakatsuji, M. Caricato, X. Li, H. P. Hratchian, A. F. Izmaylov, J. Bloino, G. Zheng and J. L. Sonnenberg, *Gaussian, Inc. Wallingford, CT*, 2009.

46   *developer.nvidia.com/cublas*.

47   T Poggio and F Girosi, *Science (80-. ).*, 1990, **24**–**7**, 978–982.

48   T. Amaral, L. M. Silva, L. A. Alexandre, C. Kandaswamy, J. M. Santos and J. M. De S, in *Proceedings - 2013 12th Mexican International Conference on Artificial Intelligence, MICAI 2013*, IEEE, 2013, pp. 114–120.

49   D. Kingma and J. Ba, *arXiv:1412.6980 [cs.LG]*, 2014, 1–15.

50   N. Srivastava, G. E. Hinton, A. Krizhevsky, I. Sutskever and R. Salakhutdinov, *J. Mach. Learn. Res.*, 2014, **15**, 1929–1958.

51   T. H. Stanley, *J. Pain Symptom Manage.*, 1992, **7**, S3-7.

52   D. J. Heal, S. L. Smith, J. Gosden and D. J. Nutt, *J. Psychopharmacol.*, 2013, **27**, 479–96.

# Supplementary information for: *"ANI-1: An extensible neural network potential with DFT accuracy at force field computational cost"*

Justin S. Smith[1], Olexandr Isayev[2,*], Adrian E. Roitberg[1,*]

[1]*Department of Chemistry, University of Florida, Gainesville, FL 32611, USA*

[2]*UNC Eshelman School of Pharmacy, University of North Carolina at Chapel Hill, Chapel Hill, NC 27599, USA*

\* Corresponding authors; email: O.I. (olexandr@olexandrisayev.com) or A.E.R. (roitberg@ufl.edu)

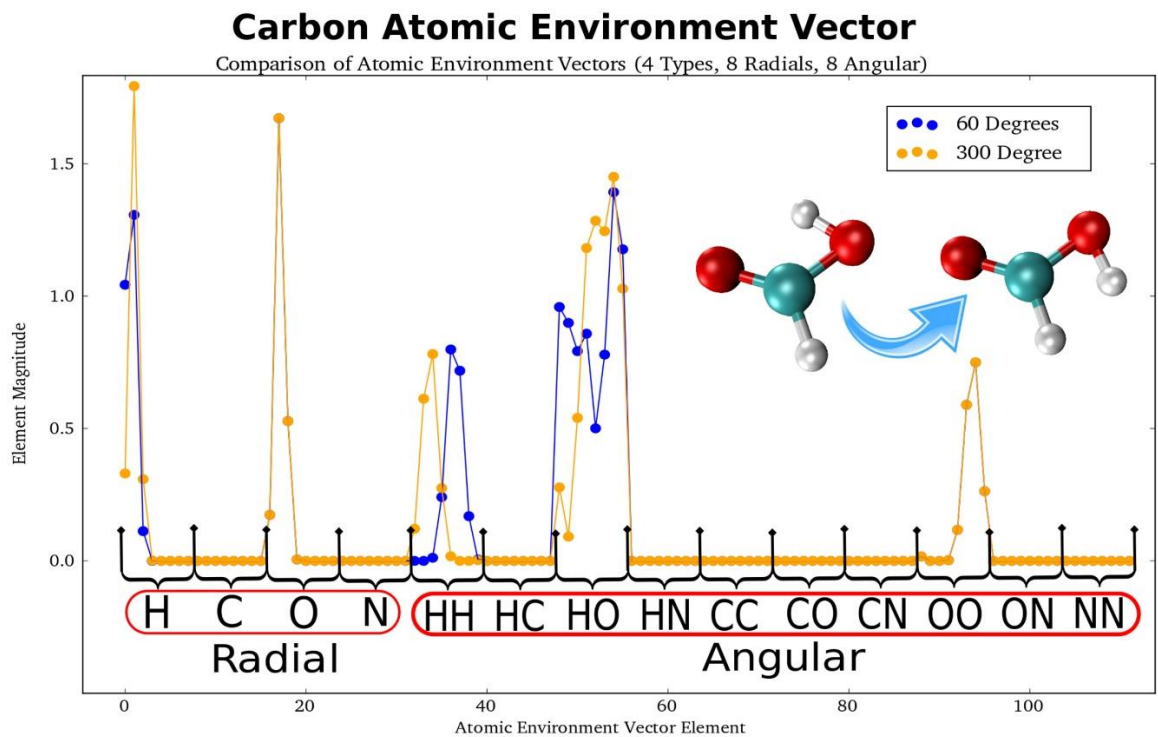

*Figure S1: A visualization of atomic environment vectors for the carbon atom ($\vec{G}_1^C$) in formic acid, computed with our modified angular symmetry functions and atomic number differentiated. The figure shows two $\vec{G}_1^C$, blue and orange, of two conformations and labels each sub-vector for clarity. The two conformation only differ in the C-O-H angle depicted in the figure.*

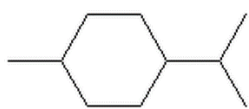 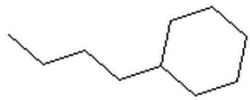 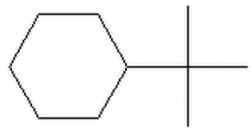 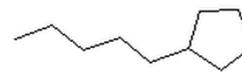

0) P-Menthane  1) N-Butylcyclohexane  2) T-Butylcyclohexane  3) Pentylcyclopentane

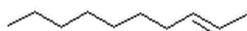 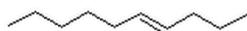 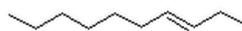 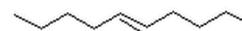

4) Trans-2-Decene  5) Trans-4-Decene  6) Trans-3-Decene  7) Trans-5-Decene

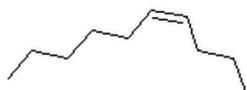 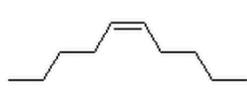 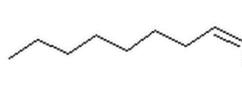 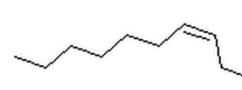

8) Cis-4-Decene  9) Cis-5-Decene  10) Cis-2-Decene  11) Cis-3-Decene

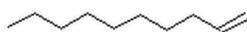

12) Dec-1-ene

*Figure S2: All structural and geometric isomers used to generate the data for the isomer case study in section 4.2. The molecular indices map to the isomer index (x-axis) of Figure 4 in Section 4.2.*

| Number of heavy atoms | Total Molecules | Max Temperature | S value | Total data points | ANI-1 test set RMSE per atom (kcal/mol/atom) |
|---|---|---|---|---|---|
| 1 | 3 | 2,000.0 | 500 | 8800 | $7.33\times10^{-2}$ |
| 2 | 13 | 1,500.0 | 450 | 39370 | $5.96\times10^{-2}$ |
| 3 | 20 | 1,000.0 | 425 | 128,880 | $4.16\times10^{-2}$ |
| 4 | 63 | 600.0 | 400 | 535,660 | $3.41\times10^{-2}$ |
| 5 | 275 | 600.0 | 200 | 1,444,890 | $3.71\times10^{-2}$ |
| 6 | 1,408 | 600.0 | 30 | 1,309,620 | $4.36\times10^{-2}$ |
| 7 | 7,850 | 600.0 | 20 | 5,276,930 | $6.65\times10^{-2}$ |
| 8 | 48,319 | 450.0 | 5 | 8,472,200 | $7.43\times10^{-2}$ |
| Total | 57,951 | - | - | 17,216,350 | $6.66\times10^{-2}$ |

*Table S1: List of information and parameters used to generate the ANI-1 data set. The first column represents the number of heavy atoms per molecule in the test set. Total represents a combination of all test sets. The molecules are obtained from the GDB-11 database.*

| Statistic (Energy units of kcal/mol) | ANI-1 Performance |
|---|---:|
| MAE | 1.316 |
| % MAE | $1.084 \times 10^{-3}$ |
| RMSE | 1.915 |
| % RMSE | $1.578 \times 10^{-3}$ |
| MAPE (%) | $4.484 \times 10^{-4}$ |
| RMSE (kcal/mol/atom) | $7.996 \times 10^{-2}$ |
| Slope | 1.000 |
| Intercept | -1.493 |
| R squared | 1.000 |
| Compute time (ms) | 286.4 |
| Data points | 8245 |
| Time per data point (µs) | 34.74 |

*Table S2: Statistics comparing the absolute energies of ANI-1 and DFT for a test set of 62 conformations of each 134 randomly selected molecules with 10 heavy atoms. Since this is a comparison of absolute energies, the range of energies is very large: from -365,343 to -243,973 kcal/mol.*

| $E^{cap}$ (kcal/mol) | RMSE | MAE | RMSE/atom | Max $|\Delta E|$ | Relative RMSE | Data points |
|---|---|---|---|---|---|---|
| \multicolumn{7}{|c|}{**134 molecules from GDB-10**} | | | | | | |
| \multicolumn{7}{|c|}{**NMS generated test set**} | | | | | | |
| 500 | 5.626 | 1.987 | 1.86E-01 | 135.966 | 5.589 | 9171 |
| 400 | 2.818 | 1.531 | 1.09E-01 | 78.449 | 2.708 | 8819 |
| 300 | 1.915 | 1.316 | 8.00E-02 | 23.876 | 1.768 | 8245 |
| 200 | 1.616 | 1.164 | 6.76E-02 | 12.722 | 1.367 | 7032 |
| 100 | 1.363 | 0.999 | 5.50E-02 | 8.226 | 0.977 | 4485 |
| 75 | 1.270 | 0.936 | 5.06E-02 | 8.226 | 0.843 | 3530 |
| 50 | 1.179 | 0.867 | 4.61E-02 | 8.226 | 0.694 | 2493 |
| 30 | 1.126 | 0.831 | 4.23E-02 | 4.551 | 0.566 | 1555 |
| 20 | 1.092 | 0.809 | 4.06E-02 | 4.332 | 0.454 | 1084 |
| 10 | 1.019 | 0.773 | 3.75E-02 | 3.953 | 0.363 | 621 |
| Min | 1.034 | 0.778 | 3.56E-02 | 3.634 | N/A | 134 |

*Table S3: The ANI-1 potentials performance on 9171 normal mode sampling (NMS) generated conformers of 134 randomly selected molecules from the GDB-10 database. $E^{cap}$ is imposed on a per molecules basis by throwing out any conformers that have energies $E^{cap}$ higher than the minimum energy for that molecule's set of conformers. This leaves only conformers closer to the minimized energy structure as $E^{cap}$ is reduced, until only the minimum energy (min) for each molecule is considered. Columns 2 through 4 show various errors to the total energies from DFT reference calculations. Column 5 shows the maximum $|\Delta E|$ over the entire data set. Column 6 shows the RMSE of energies relative to the minimum energy for each molecule's set of structures.*

| | ANI method Network performance vs data set size | | | |
|---|---|---|---|---|
| | (Error: RMSE kcal/mol) | | | |
| | Fractional Data | | Full Data | |
| Percent | Train | Valid | Test | GDB-10 Test |
| **5.00%** | 1.49 | 2.07 | 2.10 | 3.21 |
| **5.00%** | 1.56 | 2.07 | 2.13 | 3.16 |
| **5.00%** | 1.44 | 2.02 | 2.09 | 3.02 |
| **5.00%** | 1.60 | 2.06 | 2.14 | 3.11 |
| | | | | |
| **10.00%** | 1.39 | 1.73 | 1.80 | 2.68 |
| **10.00%** | 1.29 | 1.68 | 1.77 | 2.83 |
| **10.00%** | 1.44 | 1.80 | 1.83 | 2.81 |
| | | | | |
| **25.00%** | 1.18 | 1.43 | 1.45 | 2.28 |
| **25.00%** | 1.17 | 1.42 | 1.45 | 2.41 |
| **25.00%** | 1.15 | 1.40 | 1.44 | 2.46 |
| **25.00%** | 1.20 | 1.42 | 1.46 | 2.37 |
| | | | | |
| **50.00%** | 1.17 | 1.32 | 1.34 | 2.22 |
| **50.00%** | 1.20 | 1.33 | 1.36 | 2.22 |
| | | | | |
| **75.00%** | 1.09 | 1.20 | 1.21 | 2.06 |
| | | | | |
| **100.00%** | 1.16 | 1.28 | 1.28 | 1.91 |
| | | | | |
| | Baseline - No type differentiation | | | |
| **100.00%** | 3.61 | 3.78 | 3.84 | 6.55 |
| | | | | |
| | Baseline – CM/MLP | | | |
| **5.00%** | 42.17 | 46.61 | 48.07 | 1047.84 |
| **10.00%** | 45.49 | 45.77 | 47.14 | 1457.68 |
| **25.00%** | 35.44 | 38.03 | 38.15 | 503.57 |
| **50.00%** | 35.33 | 39.28 | 38.63 | 1422.11 |
| **75.00%** | 34.56 | 36.61 | 36.71 | 460.87 |
| **100.00%** | 33.79 | 35.96 | 36.09 | 493.70 |

*Table S4: Shows how the ANAKIN-ME method scales with the size of the training set as well as information about two baseline methods trained on the same data set. The "Percent" column shows what percentage of the 17.2 million data points was used to train, validate, and test the model. The train and validate columns show the RMSE of the actual training and validation set, fractional data, used to train the model while the test sets are always full sets. The first baseline method shows how the ANAKIN-ME method performs without differentiating atomic numbers within the AEVs. The second baseline shows the performance of a sorted coulomb matrix with a multilayer perceptron (CM/MLP) neural network model on the ANI-1 data set with training set size scaling.*